\documentclass[aip,apl,reprint]{revtex4-1}
\usepackage{graphicx,graphics,epsfig,color}
\usepackage{epstopdf}
\usepackage{amsmath,amsmath}
\usepackage{dcolumn}
\usepackage{bm}
\usepackage{hyperref}

\newcommand{\mr}[1]{{\mathrm{#1}}} 		

\begin{document}


\title[Single Atom Imaging with an sCMOS camera]{Single Atom Imaging with an sCMOS camera}

\author{C.J. Picken}  
\affiliation{Department of Physics, University of Strathclyde, 107 Rottenrow East, Glasgow, G4 0NG, UK}%
\author{R. Legaie} 
\affiliation{Department of Physics, University of Strathclyde, 107 Rottenrow East, Glasgow, G4 0NG, UK}%
\author{J. D. Pritchard}  \email{jonathan.pritchard@strath.ac.uk}
\affiliation{Department of Physics, University of Strathclyde, 107 Rottenrow East, Glasgow, G4 0NG, UK}%

\date{\today}

\begin{abstract}

Single atom imaging requires discrimination of weak photon count events above background and has typically been performed using either EMCCD cameras, photomultiplier tubes or single photon counting modules. sCMOS provides a cost effective and highly scalable alternative to other single atom imaging technologies, offering fast readout and larger sensor dimensions. We demonstrate single atom resolved imaging of two site-addressable optical traps separated by 10~$\mu$m using an sCMOS camera, offering a competitive signal-to-noise ratio at intermediate count rates to allow high fidelity readout discrimination (error $<10^{-6}$) and sub-$\mu$m spatial resolution for applications in quantum technologies. 
\end{abstract}


\maketitle

\section{Introduction}

Recent developments in quantum information processing have lead to a requirement for resolved single atom imaging of isolated atomic qubits in microscopic optical traps \cite{sortais07,urban09}, ion traps \cite{streed11} or optical lattices where quantum gas microscopes offer a route to quantum simulation \cite{bakr09,haller15}.

Single atom imaging requires both spatial and number resolution, where a finite number of scattered photons are collected by high numerical aperture (NA) optics in order to obtain a large collection efficiency from atoms in microscopic traps\cite{alt02,sortais07,pritchard16,lester14}, optical lattices\cite{sherson10} or magnetic traps \cite{leung14}. This enables multiple readouts of the same atom with hyperfine resolved detection of atomic qubits \cite{kwon17} or counting of individual atoms in ensembles of over 100 atoms\cite{hume13}.

Due to the low photon numbers reaching the detector (typically $\sim10~\mr{photons}$/ms/atom), single atom detection has typically been performed by single photon counting modules (SPCMs)\cite{schlosser01} or photomultiplier tubes (PMTs)\cite{nagourney83} offering extremely low dark counts but only a single pixel. Therefore spatially resolved detection has until now been exclusively performed with electron-multiplying charge-coupled device (EMCCD) cameras\cite{grunzweig10} or a standard scientific CCD coupled with an intensifier\cite{fuhrmanek10}.

In this paper we present single atom number resolved measurements using a scientific complementary metal-oxide semiconductor (sCMOS) camera. Unlike an (EM)CCD camera, each pixel is read out independently, removing clock induced charge noise and permitting higher readout speeds whilst offering a superior signal-to-noise-ratio (SNR) for intermediate incident photon rates. SCMOS cameras are thus an attractive candidate for scalable quantum information processing (QIP), providing  larger sensor sizes and the ability to perform high-speed real time single pixel processing using on-board field-programmable gate array (FPGA) hardware\cite{zyla5_5}.

\section{Camera SNR}
Quintessential to performing imaging with single atom resolution is overcoming detector noise to discriminate the weak photon events from a single atom over the background count rate. The signal-to-noise ratio (SNR) of a camera is given by \cite{fullerton12}
\begin{equation}
\mr{SNR}=\frac{nQE}{\sqrt{F_n\!^2QE(n+n_b)+(\delta_{ro}/M)^2}},
\end{equation}
where $n$ is the number of incident photons per pixel, $n_b$ is the number of background photons per pixel, $QE$ is the quantum efficiency, $F_n$ the noise factor, $M$ the multiplication factor and $\delta_\mr{ro}$  is the camera readout noise. Other camera noise factors such as clock induced charge and dark noise have been considered negligible for the cameras and imaging timescales examined in this paper. From the above relation it can be seen that having a low readout noise coupled with a large $QE$ is crucial to achieve high SNRs at low photon levels.

Standard scientific CCD detectors perform with SNRs close to that of an ideal detector for high photon numbers due to their near perfect noise factor, $F_n=1$. However in the limit of few photons $\approx 10~\mr{photons}$/px, the SNR suffers due the high readout noise\cite{orcaR2} $\delta_{ro}>6e^-$. An EMCCD camera overcomes this constraint through an electron multiplying process which amplifies the signal up to $M\sim1000$, allowing an effective readout noise $\delta_{ro}/\rm{M}<~1e^-$ to be achieved\cite{andoriXon}. This multiplication process results in an increased noise factor, $F_n=\sqrt{2}$, but makes it an incredibly powerful tool for low photon imaging applications, such as imaging single atoms and ions.

Recent advances in sCMOS cameras have made it a contender in low light imaging. Each pixel is read out independently, enabling larger sensor sizes with a high speed FPGA to process readout\cite{krishnaswami14,ma13}. The use of ultra low noise MOSFETs reduces the readout noise to values\cite{zyla5_5} as low as $2e^-$. This ensures fast integrated readout times and since there is no additional amplification process a near perfect noise factor $F_n=1$ is achieved, allowing the sCMOS to be competitive at intermediate photon levels of 10-100 $\mr{photons}$/px. 

The sCMOS camera used in this paper is the Andor Zyla 5.5. In global shutter mode with 200~MHz readout speed, $\delta_{ro}=2.2~e^-$ rms with a maximum QE of 60~\%.
Table~1 compares the SNRs for top of the line cameras using EMCCD and CCD technology as well as the Zyla. It can be seen when operating in the range of the best QE that the Zyla outperforms the other technologies for mid level photon events while still providing a competitive SNR for low level imaging. At our imaging wavelength, $\lambda=852$~nm, $QE=22$~\% giving a performance that is comparable to an EMCCD for 100~$\mr{photons}$/px. 

\begin{table}[h!]
\begin{ruledtabular}
\begin{tabular}{c c c c c c c}
Detector & \multicolumn{2}{c}{10~$\mr{photons}$/px}& \multicolumn{2}{c}{100~$\mr{photons}$/px} &   \multicolumn{2}{c}{1000~$\mr{photons}$/px}   \\
& Best&852 nm & Best & 852 nm & Best  & 852 nm\\
\hline
Zyla & 1.8 & 0.8 & 7.4 & 4.2 & 24.4 & 14.7 \\
EMCCD & 2.1 & 1.7 & 6.7 & 5.2 & 21.2 & 16.6  \\
CCD &0.7 & 0.2 & 5.4 & 1.8 & 24.7 & 11.5\\
\end{tabular}
\end{ruledtabular}
\caption{SNR comparison of different available camera technologies for both the best QE and the QE at 852 nm. The EMCCD considered is the Andor iXON Ultra 897 ($QE=90~\%,QE_{852}=60~\%,\delta_{ro}=89$e$^- $(17~MHz operation)$,M=1000,F_n=\sqrt{2}$)\cite{andoriXon} and the CCD is the Hamamatsu Orca-R2 ($QE=70~\%,QE_{852}=20~\%,\delta_{ro}=10$e$^- $(fast scan mode)$,F_n=1$)\cite{orcaR2}.
\label{ATOM}}
\end{table}

\begin{figure}[t]
\includegraphics{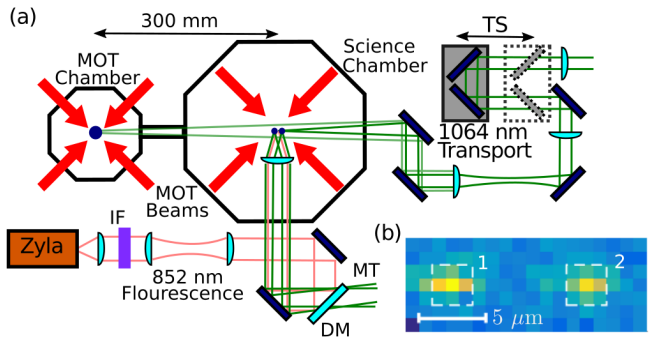}
\caption{(Color online). (a) Single atom imaging setup. IF= Interference Filter, DM= Dichroic Mirror, MT= Microtrapping Light and TS= Translation Stage (b) Single shot image of two single atoms separated by~10~$\mu$m\label{fig1}}
\end{figure}

\section{Setup}

A schematic of the setup used to load single atoms can be seen in Fig.~\ref{fig1}(a). Preparation begins in the MOT chamber located 300~mm away from the atom trapping site. We load 10$^6$ caesium atoms in a 3D magneto-optical trap (MOT) in 1~s before transferring them into an optical dipole trap with wavelength $\lambda=$1064~nm, a beam waist $w_0=43$~$\mu$m and trap depth $U_0=600~\mu$K where 10~\% loading efficiency is achieved. After a polarization gradient cooling stage\cite{dalibard89} to cool atoms to 10~$\mu$K, optical transport between the chambers is achieved using a motorized translation stage (Thorlabs DDS220/M) in 800~ms. Following successful transport to the science chamber the dipole trap is overlapped with a pair of microscopic tweezer traps for a period of 60~ms with weak 3D cooling light,~$I=I_\mathrm{sat}$ per beam with a detuning $\Delta=-6\Gamma$ to load multiple atoms into each of the microscopic traps, where $I_\mathrm{sat}=2.7$~mW/cm$^2$ is the polarization-averaged saturation intensity and $\Gamma/2\pi=5.22$~MHz is the spontaneous decay rate of the transition\cite{steckCs}.

The microscopic tweezer traps are formed using a diffraction-limited aspheric lens with ${NA}=0.45$ (Geltech 355561) \cite{sortais07} mounted in vacuum, providing a large collection efficiency for the light emitted by each atom~$\sim$~5.4~\%. In order to suppress background electric fields due to the close proximity of the atoms to the lens surface, the lens is coated in a layer of indium tin oxide (ITO) which reduces the transmission to 79~\%. The traps have a waist of 1.95~$\mu$m, which at $\lambda=1064$~nm and 28~mW power results in a trap depth of $U_0=1.2$~mK with radial (axial) trap frequencies of $\nu_{r(z)}=47~(5.7)$~kHz. 

After loading atoms into the microscopic dipole traps, a 100~ms single atom loading stage is performed at a detuning of $\Delta=-8\Gamma$ and an intensity of $\sim$0.5~$I_\mathrm{sat}$ per beam. To remove light shifts associated with the trapping potential, cooling light is chopped out of phase with the trapping light at 1~MHz with a 35~\% duty cycle. Due to the small trapping volume single atom loading via collisional blockade \cite{schlosser02} is achieved, where light-assisted collisions (LACs) cause pairs of atoms to be lost due to excitation of unstable molecular potentials, resulting in probabilistic loading of either 0 or 1 atom in each trap. Using a release-recapture method\cite{tuchendler08} we measure the temperature of the single atoms to be 15~$\mu$K after this stage.

\section{Single Atom Imaging}
Light scattered by the atoms is collected by the aspheric lens and separated from the microtrap beams using a dichroic mirror as shown in Fig.~\ref{fig1}(a). The collected light is then focused by a $f=200$~mm lenses to create a confocal imaging setup where the beam is imaged onto the Zyla chip through a relay telescope ($f_1=100$~mm, $f_2=30$~mm) to enable filtering in the Fourier plane using narrowband interference filters to block the 1064~nm light reaching the camera and transmit the 852~nm light from the atoms, with a measured transmission of 83~\%. The combined detection efficiency of the imaging system including the filters is 3.5~\%. From Zemax calculations we obtain a paraxial magnification of -20.5 at the intermediate focus of the 200~mm lens, resulting in a total magnification of M=+0.62 between object plane in the chamber and the image plane on the Zyla; this corresponds to an effective pixel size of 1~$\mu$m. Calibration of the relay imaging using a USAF 1951 resolution test chart finds a sub-pixel point spread function of $0.7 \pm 0.1$~$\mu$m. 

In order to detect single atoms, a sufficient number of photons must be scattered in order to distinguish between scattered background light and the events due to the single atom. The photon scattering rate for a single atom is given by\cite{foot05}
\begin{equation}
\Gamma_\mr{sc}=\frac{\Gamma}{2}\frac{I/I_\mathrm{sat}}{1+I/I_\mathrm{sat}+4(\Delta/\Gamma)^2},
\end{equation}
where $\Delta$ it the detuning and $I$ the intensity. The above relation clearly shows that the largest scattering rate is achieved on resonance, however imaging on resonance causes heating and eventually the loss of the atom from the trap. For our experiments imaging is performed with 0.5~$I_\mathrm{sat}$ of cooling power per beam at a detuning of $\Delta=-3~\Gamma$, utilising the same out of phase chopped light pulses described above resulting in an effective photon scattering rate of $\Gamma_\mr{sc}=450~\mr{photons}$/ms, with an expected flux of $15~\mr{photons}$/ms incident on the camera. Fig.~\ref{fig1}(b) show a typical image obtained imaging in this way using a total imaging time of 40~ms, showing two clearly resolved optical traps with a separation of 10~$\mu$m.

\begin{figure}[t]
\includegraphics{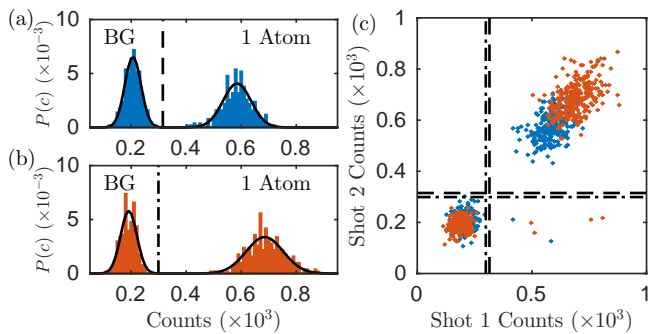}
\caption{(Color online). Probability distribution of the counts from 500 measurements recorded for (a) trap 1 and (b) trap 2 using a 40 ms exposure. (c) Correlation plot showing counts from the first image plotted against the counts for an image taken 50 ms later\label{fig2}}
\end{figure}
Single atom loading of the trap sites is verified through the emergence of a bimodal probability distribution for the number of counts detected within a $3\times3$ pixel region of interest centred on each trap following the LAC stage. Figure~\ref{fig2}(a,b) shows the probability distribution obtained from 500 repeated measurements using the imaging parameters described above for each trap, which clearly reveals two well separated distributions corresponding to a Poisson distribution centred at a mean count rate $\mu_0$ when no atom is loaded, and a second distribution centred at a mean $\mu_1$ when an atom is present. The reduced count rate observed from atoms in trap 1 is due to a weak standing wave in the retro-reflected MOT beams creating a position-sensitive scattering rate.


Further evidence for single atom loading is obtained by comparing the results of imaging the same trap twice in a single measurement run. Figure~\ref{fig2}(c) shows correlations between counts in shot 1 and counts in a second shot taken 50~ms later, revealing two distinct clusters associated with having 0 (1) atoms present in both shots, and a small number of points in the lower right quadrant corresponding to an atom initially loaded in shot 1 but having been lost by shot 2. Collating the data in this way also clarifies the ability to retain the atom after readout, with $>$~98~\% retention probability for having an atom present in the second shot for both traps (dominated by collisional loss from background gas as shown below). With the LAC stage, we never observe counts corresponding to double load events confirming a robust single atom loading sequence.

In order to analyse the data, we approximate the Poisson count distributions with a large mean to a bimodal Gaussian distribution using the equation
\begin{equation}\label{eq:P}
P(c)=p_0G(c,\mu_0,\sigma_0)+p_1G(c,\mu_1,\sigma_1),
\end{equation}
where $p_{i}$ is the probability of loading zero or one atom and $G(c,\mu,\sigma)=1/\sqrt{2\pi\sigma^2}\exp\{-(c-\mu)^2/(2\sigma^2)\}$ is a normalised Gaussian distribution. Fitting the data in Fig.~\ref{fig2} we obtain parameters summarised in Tab.~\ref{tab:fit}, with both traps loading atoms $>50$~\% of the time corresponding to sub-Poissonian loading as observed in other experiments \cite{schlosser01}. 

\begin{table}[b]
\begin{ruledtabular}
\begin{tabular}{c c c c c c}
Trap & \multicolumn{1}{c}{$\mu_0$}& \multicolumn{1}{c}{$\sigma_0$} &   \multicolumn{1}{c}{$\mu_1$} &   \multicolumn{1}{c}{$\sigma_1$}  &   \multicolumn{1}{c}{$p_1 (\%)$}    \\
\hline
1 & 206 & 29 & 586 & 51 &  52\\
2 & 191 & 29 & 685 & 67  & 57\\
\end{tabular}
\end{ruledtabular}
\caption{Fit parameters for data in Fig.~2 to Eq.~\ref{eq:P}.\label{tab:fit}}
\end{table}

For each measurement the count rate in the first shot is used to determine if an atom is present in the trap by introducing a threshold value $c_\mr{min}$ above which an atom is assigned to the trap. Data in shot two is then analysed conditional upon detection in shot 1, either through fitting the resulting bimodal distribution to extract $p_1$ or again using a threshold method. The error $\mathcal{\epsilon}$ associated with correctly labelling an atom in the trap is calculated using
\begin{equation}
\mathcal{\epsilon} = \displaystyle\int_{c_\mathrm{min}}^\infty p_0G(c,\mu_0,\sigma_0)\mr{d}c = \frac{p_0}{2}\left[1-\mathrm{erf}\left(\frac{c_\mathrm{min}-\mu_0}{\sqrt{2}\sigma_0}\right)\right],
\end{equation}
whilst the acceptance $\mathcal{A}$ (defined as the fraction of single atom load events accepted using $c>c_\mr{min}$) is given by
\begin{equation}
\mathcal{A} = \displaystyle\int_{c_\mathrm{min}}^\infty G(c,\mu_1,\sigma_1)\mr{d}c= \frac{1}{2}\left[1-\mathrm{erf}\left(\frac{c_\mathrm{min}-\mu_1}{\sqrt{2}\sigma_1}\right)\right].
\end{equation}

\begin{figure}[t]
\includegraphics{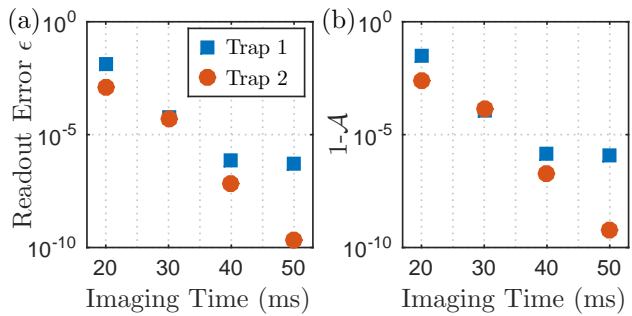}
\caption{(Color online). (a) Readout error and (b) Acceptance as a function of imaging duration.}
\label{fig3}
\end{figure}

Figure~2 shows cut-off values chosen to minimise the overlap volume between the two probability distributions corresponding to $c_\mathrm{min}=346,345$ for trap 1 and 2 respectively. The corresponding error is $\mathcal{\epsilon}<8\times10^{-7}$ with an acceptance~$\mathcal{A}>99.99\%$ for both traps, corresponding to high measurement fidelity. Figure~\ref{fig3} shows the evolution of $\mathcal{\epsilon}$ and $\mathcal{A}$ as a function of imaging duration for both traps, showing that 40~ms provides an optimal readout time as for longer imaging durations heating in trap 1 limits readout fidelity.

\begin{figure}[t]
\includegraphics{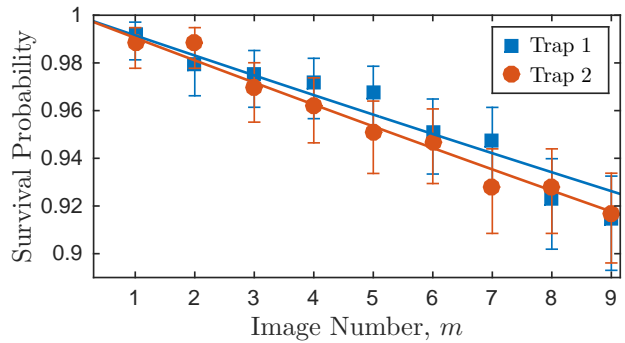}
\caption{(Color online). Single atom retention for multiple 40~ms measurements at -3$\Gamma$ separated by 60~ms. The single shot retention $\alpha$, is found to be~$>$~99~\% for both traps when fitted to $P(m)=\alpha^m$.}
\label{fig4}
\end{figure}

Finally, for application in state readout, we require that not only can we detect the presence or absence of an atom with high fidelity, but that the imaging process is non-destructive to avoid losing the atom from the trap corresponding to a high retention rate. To accurately determine the retention between measurements, we perform 10 sequential imaging sequences each separated by 60~ms. Following the 40~ms imaging pulse, the atoms are heated to 25~$\mu$K, and a short 10~ms cooling cycle is added between images to maximise retention. It can be seen from Fig.~\ref{fig4} that we can reliably retain the atom for many sequential images, from which we obtain a single shot retention $\alpha>99$~\% in both traps from fitting data to $P=\alpha^m$, where $m$ is the number of images and $P$ is the survival probability. For comparison, measurement of the trap lifetime returns a 1/$e$ lifetime exceeding 6~s for both traps and a survival probability of 87~\% for trap~1 and 90~\% for trap~2 after 600~ms. These results indicate that the main limitation in our measurements is the background-limited lifetime of the atoms in the trap, which we estimate to correspond to a pressure $P\lesssim3\times10^{-9}$~Torr \cite{arpornthip12}.

\section{Comparison to EMCCD hardware}
The results above demonstrate that high fidelity state detection is achievable using sCMOS sensors despite the limited quantum efficiency compared to EMCCD cameras as summarised in Table~1, meaning simply scattering more photons and thus slightly increased heating to achieve the same number of photon detection events but still providing excellent performance. For experiments requiring fast repetition rate, readout time is also a consideration. Using the present hardware we are able to perform multiple images with a minimum separation of 10~ms for frame transfer, compared to 100~ms for a recent demonstration of non-destructive quantum state readout with an EMCCD camera\cite{kwon17}, significantly reducing sensitivity to losses arising from background-gas collisions between detection events and enabling faster experiment cycle times. Another important factor however is cost, with the Andor Zyla providing comparable performance for less than a third of the cost of the popular Andor iXon EMCCD camera at time of writing, making sCMOS highly competitive for single atom imaging applications.

To extend the current results to enable a hyperfine-resolved imaging for quantum state readout, it is necessary to scatter light from the upper hyperfine ground-state using a closed-transition and collect a sufficient number of photons to discriminate the counts above background before the atom undergoes a hyperfine changing transition due to off-resonant Raman processes or imperfect polarisation of light. Kwon \emph{et al.} have demonstrated this for Rb, where with 2~\% conversion from photons to counts they observed non-destructive quantum state discrimination from 7500 photon scattering events\cite{kwon17}. In the present Cs experiment we obtain an equivalent conversion efficiency of 2.1, 2.7 \% for the two traps respectively, meaning similar results should be achievable. However, our current setup would need modification to achieve the required control of polarisation of both trapping and imaging light necessary to minimise hyperfine state depumping during imaging to demonstrate this. Switching to Rb would be even more favourable, as the increase in $QE$ at 780~nm would enable detection efficiencies of $\sim3.5~\%$.

Finally, due to the FPGA hardware integrated alongside the sCMOS sensor chip\cite{zyla5_5} it should be possible to perform high-speed single pixel readout and image processing on the camera itself\cite{ma13}, resulting in high-speed state detection and removing the need for frame transfer. Progress towards this goal is currently limited by the proprietary camera firmware, however in future customisable hardware will become more widely available.

\section{Conclusion}
We have demonstrated resolved single atom imaging with a sCMOS camera, with the ability to perform multiple non-destructing measurements with high fidelity single atom detection ($\mathcal{\epsilon}<8\times10^{-7}$) and a retention $>99$~\% in two spatially resolved optical traps. Despite the limited $QE$ of the camera at the imaging wavelength, we achieve comparable performance to experiments using costly EMCCD based detectors with a superior SNR possible at the intermediate photon count rates. This technology offers a viable, cost-effective alternative to other currently used techniques in low photon detection, and has the additional benefits of a larger sensor size and the ability to independently readout single pixels with the high-speed integrated FPGA hardware for performing scalable quantum state detection.

\begin{acknowledgments} 
We thank Mark Saffman for his valuable assistance in system design and to Aidan Arnold and Paul Griffin for careful reading of the manuscript. This work was supported by funding from the ESPRC (Grant No. EP/N003527/1). The data presented in the paper are available\cite{picken17data}.
\end{acknowledgments}


%

\end{document}